\documentclass[10pt,letterpaper,twocolumn]{article} 
\usepackage{ol2}
\usepackage{amsmath,amssymb}
\usepackage{psfig,bm}

\begin{document}

\twocolumn[

\title{Correct Definition of the Poynting Vector in Electrically and
  Magnetically Polarizable Medium Reveals that Negative Refraction is
  Impossible}

\author{Vadim A. Markel}
\affiliation{Departments of Radiology and Bioengineering, University
  of Pennsylvania, Philadelphia, PA 19104\\
{\tt vmarkel@mail.med.upenn.edu}}

\begin{abstract}
  I compute from first principles the local heating rate $q$ (the
  amount of electromagnetic energy converted to heat per unit time per
  unit volume) for electromagnetic waves propagating in magnetically
  and electrically polarizable media. I find that, in magnetic media,
  this rate has two separate contributions, $q^{(V)}$ and $q^{(S)}$,
  the first coming from the volume of the medium and the second from
  its surface. I argue that the second law of thermodynamics requires
  that the volume contribution be positive and that this requirement,
  in turn, prohibits negative refraction. This result holds for active
  or passive media and in the presence of anisotropy and spatial
  dispersion.
\end{abstract}
\date{\today}
]
\section{Introduction}
\label{sec:intro}

Macroscopic electromagnetic theory of material media which can
simultaneously support electric and magnetic polarizations denoted by
${\bf P}$ and ${\bf M}$, respectively, has been developed over a
century ago and is exposed in many standard textbooks. However, in the
optical frequency range and at higher frequencies, this theory has
long been viewed as purely abstract and nonempirical. Even at much
lower frequencies, materials which simultaneously exhibit nonzero
magnetic and electric susceptibilities (and are sufficiently
transparent to allow any noticeable penetration of electromagnetic
field into their interior) are quite rare and exotic.

While it is possible to argue about the physical attainability of
artificial materials with nonzero electric and magnetic
susceptibilities in any given frequency range, nothing precludes us
from formally developing the electrodynamics of such media based on
the macroscopic Maxwell equations. In particular, this approach was
adopted by Veselago in the now famous paper
(Ref.~\onlinecite{veselago_68_1}). Veselago was interested in
materials whose electric permittivity $\epsilon$ and magnetic
permeability $\mu$ are simultaneously negative and which can exhibit
the so-called negative refraction - the physical effect which takes
place when an electromagnetic wave entering the medium (e.g., from
vacuum) is refracted at the ``negative'' Snell's angle.

After the publication of a more recent paper by Pendry in which a
perfect (subwavelength-focusing) lens built from a
negatively-refracting material was proposed~\cite{pendry_00_1},
enormous attention was attracted to negative refraction. Numerous
proposals for manufacturing artificial materials with negative
refraction have been put forth.  There has also been a rigorous effort
to demonstrate negative refraction experimentally; see, for example,
Refs.~\onlinecite{ramakrishna_05_1,soukoulis_07_1,shalaev_07_1} and
references therein.

Simultaneously with the activities mentioned above, there has also
been a persistent effort to subject the physical attainability of
negative refraction to doubt. Perhaps, the most consequential of such
exploits is the recent paper by Stockman~\cite{stockman_07_1} in which
it is shown from the causality principle that, in a
negatively-refracting material, the rate of dissipation of the
electromagnetic energy into heat can not be lower than a certain
threshold and that any attempt to compensate for such dissipation,
e.g., by introducing optical gain, will necessarily destroy the
negative refraction. It is worthwhile to note that low dissipative
losses are essential for the realization of the original Pendry's
proposal for the perfect lens.

In this article I confront the phenomenon of negative refraction with
another fundamental physical principle, the second law of
thermodynamics. I show that the general requirement for negative
refraction in isotropic media, namely,

\begin{equation}
\label{NR_cond}
{\rm Im}(\epsilon \mu) < 0
\end{equation}

\noindent
is in contradiction with the latter. To do so, I compute the heating
rate $q({\bf r})$ in a magnetically and electrically polarizable
medium. I do this by two methods, one involving the expression
$-\nabla \cdot {\bf S}$, where ${\bf S}$ is the Poynting vector, and
the other involving the expression ${\bf J}\cdot {\bf E}$, where ${\bf
  E}$ is the electric field and

\begin{equation}
\label{J_def}
{\bf J} = \frac{\partial {\bf P}}{\partial t} + c \nabla \times {\bf M}
\end{equation}

\noindent
is the total current induced in the medium (we assume that there are
no {\em external} currents or charges). Quite unexpectedly, I obtain
different results. I claim that the explanation for this discrepancy
is that the Poynting vector in a magnetically polarizable medium must
be defined by

\begin{equation}
\label{S_B}
{\bf S} = \frac{c}{4\pi} {\bf E} \times {\bf B}
\end{equation}

\noindent
rather than by the commonly used formula

\begin{equation}
\label{S_H}
{\bf S} = \frac{c}{4\pi} {\bf E} \times {\bf H} \ .
\end{equation}

\noindent
Arguments for the validity of (\ref{S_B}) are given below.

When the definition (\ref{S_B}) is adopted, the two methods of
computing $q({\bf r})$ give the same result. It further turns out that
the heating rate has two separate contributions: one coming from the
volume and the other from the surface of the medium. These
contributions are denoted by $q^{(V)}$ and $q^{(S)}$ below. The total
(that is, integral over the body volume) heat absorbed per unit time
is given by the formula

\begin{equation}
\label{W_def}
Q = \int_V q^{(V)}({\bf r}) d^3 r + \oint_S q^{(S)}({\bf r})d^2 r \ ,
\end{equation}

\noindent
where the first integral is evaluated over the body volume and the
second over the surface. The quantity $Q$ computed according to
(\ref{W_def}) is exactly the same as in the conventional theory.
However, my calculations show that the volume contribution, $q^{(V)}
\propto {\rm Im}(\mu \epsilon)$. I argue that in passive media, the
second law of thermodynamics requires that $q^{(V)}>0$ in
contradiction with the inequality (\ref{NR_cond}). In optically active
media, it is possible to have $q^{(V)}<0$ but the condition for
negative refraction is then reversed and reads ${\rm Im}(\mu\epsilon)
<0$. Thus I come to the conclusion that negative refraction is not
possible in either passive or active media.

The paper is organized as follows. In Section~\ref{sec:Q}, I compute
the volume contribution to the heating rate, $q^{(V)}$, for a
monochromatic plane wave by two different methods and obtain two
different expressions. In Section~\ref{sec:Poynting}, I argue that the
reason for this discrepancy is incorrect definition of the Poynting
vector ${\bf S}$.  When the correct definition (\ref{S_B}) is adopted,
the two methods yield the same result. Also, in
Section~\ref{sec:Poynting}, $q^{(V)}$ is computed for general
monochromatic fields (not necessarily plane waves). In
Section~\ref{sec:surface} I compute the surface contribution to the
heating rate, $q^{(S)}$. Also in this section, the zero-frequency limit
is discussed. In Section~\ref{sec:thermo} I give a detailed proof that
the second law of thermodynamics requires that $q^{(V)}>0$. In
Section~\ref{sec:causality}, I discuss compatibility of the obtained
expressions for the heating rate with the causality principle. In
Section~\ref{sec:anisotropy}, I show that negative refraction is not
possible even in anisotropic and nonlocal media.  Finally,
Sections~\ref{sec:discussion} and \ref{sec:conclusions} contain a
discussion and a summary of obtained results.

\section{Computation of the Heating Rate}
\label{sec:Q}

The heating rate $q$ is defined as the energy absorbed and transformed
into heat by a material per unit volume (or surface, if there is a
surface contribution), per unit time. In the case of oscillating
electromagnetic fields, this energy must be averaged over time periods
which are much larger than the characteristic period of oscillations.
In this section, I use two different methods to compute $q$ for a
monochromatic plane wave propagating in a homogeneous, isotropic
medium characterized by scalar and local (but time-dispersive)
functions $\epsilon(\omega) = \epsilon^{\prime}(\omega) + i
\epsilon^{\prime\prime}(\omega)$ and $\mu(\omega) =
\mu^{\prime}(\omega) + i\mu^{\prime\prime}(\omega)$.

\subsection{First derivation of $q$}
\label{subsec:first}

First, we use the well-known formula for $q$ which can be found in
many standard textbooks, namely,

\begin{equation}
\label{Q_Landau}
q^{\rm (conv)} = \frac{1}{4\pi}\left\langle {\bf E} \cdot \frac{\partial {\bf D}}{\partial t} + {\bf H} \cdot \frac{\partial {\bf B}}{\partial t} \right \rangle \ ,
\end{equation}

\noindent
where $\langle \ldots \rangle$ denotes time averaging and the
quantities ${\bf E}, {\bf D}$ and ${\bf H}, {\bf B}$ are the electric
field and displacement and the magnetic field and induction,
respectively. The superscript ``${\rm (conv)}$'' has been used to
indicate that (\ref{Q_Landau}) gives the conventional result for the
heating rate. In the case of a monochromatic field of frequency
$\omega$, (\ref{Q_Landau}) can also be written as

\begin{equation}
\label{Q_Landau_monochrom}
q^{\rm (conv)} = \frac{\omega}{4\pi}\left[ \epsilon^{\prime\prime}(\omega) \langle {\bf
    E}^2\rangle + \mu^{\prime\prime}(\omega) \langle {\bf H}^2 \rangle
\right] \ .
\end{equation}

\noindent
Note that (\ref{Q_Landau}),(\ref{Q_Landau_monochrom}) are quadratic in
electromagnetic fields; correspondingly, ${\bf E}, {\bf D}, {\bf H}$
and ${\bf B}$ are defined in these expressions as real-valued
quantities.

Let us take one step further and evaluate (\ref{Q_Landau_monochrom})
for a plane wave propagating in a homogeneous medium. We shall seek an
expression for the heating rate which contains only the amplitude of
the electric, but not of the magnetic, field. To this end, we write

\begin{equation}
\label{EH_real_complex}
{\bf E} = {\rm Re}[{\bf E}_0 e^{i({\bf k} \cdot {\bf r} - \omega t)}]
\ , \ \
{\bf H} = {\rm Re}[{\bf H}_0 e^{i({\bf k} \cdot {\bf r} - \omega t)}] \ ,
\end{equation}

\noindent
where ${\bf E}_0$ and ${\bf H}_0$ are complex field amplitudes.  Time
averaging yields $\langle {\bf E}^2\rangle = (1/2)\vert {\bf E}_0
\vert^2 \exp(-2{\bf k}^{\prime\prime} \cdot {\bf r})$ and analogously
for the magnetic field. Here ${\bf k}^{\prime\prime} = {\rm Im}({\bf
  k})$ and ${\bf k}$ satisfies ${\bf k} \cdot {\bf k} = \mu\epsilon
(\omega/c)^2$. We now substitute the expressions for the time averages
$\langle {\bf E}^2 \rangle$ and $\langle {\bf H}^2 \rangle$ in terms
of the field amplitudes ${\bf E}_0$ and ${\bf H}_0$ into
(\ref{Q_Landau_monochrom}) to obtain

\begin{equation}
\label{Q_Landau_monochrom_planewave}
q^{\rm (conv)} = \frac{\omega}{8\pi}\left[\epsilon^{\prime\prime}(\omega) \vert {\bf
    E}_0 \vert^2 + \mu^{\prime\prime}(\omega) \vert {\bf H}_0 \vert^2
\right] e^{-2 {\bf k}^{\prime\prime} \cdot {\bf r} } \ .
\end{equation}

\noindent
Further, we want to express $\vert {\bf H}_0 \vert^2$ in terms of
$\vert {\bf E}_0 \vert^2$. From the Maxwell equation $c \nabla \times
{\bf E} = - \partial {\bf B} / \partial t$ and from ${\bf B}_0 = \mu
{\bf H}_0$, it follows that ${\bf H}_0 = [c/\omega \mu(\omega)] {\bf
  k} \times {\bf E}_0$. Therefore, $\vert {\bf H}_0 \vert^2 = (c /
\omega \vert \mu \vert)^2 ({\bf k} \times {\bf E}_0) \cdot ({\bf k}^*
\times {\bf E}_0^*) = (c / \omega \vert \mu \vert)^2 [({\bf k} \cdot
{\bf k}^*)({\bf E}_0 \cdot {\bf E}_0^*) - ({\bf k} \cdot {\bf
  E}_0^*)({\bf k}^* \cdot {\bf E}_0)]$. The wave vector of a
propagating (that is, not evanescent) wave can always be written as
${\bf k} = k \hat{\bf u}$, where $\hat{\bf u}$ is a purely real unit
vector such that $\hat{\bf u} \cdot \hat{\bf u} = 1$ and $k^2 =
\mu\epsilon (\omega/c)^2$ is a complex scalar. In this case, ${\bf k}
\cdot {\bf k}^* = \vert \mu(\omega) \epsilon(\omega) \vert
(\omega/c)^2$ and ${\bf k} \cdot {\bf E}_0^* = {\bf k}^* \cdot {\bf
  E}_0 = 0$. The final expression for the heating rate then becomes

\begin{equation}
\label{Q1}
q^{\rm (conv)} = \frac{\omega}{8\pi}
\left[ \epsilon^{\prime\prime}(\omega) + \frac{\vert \epsilon(\omega) \vert}{\vert \mu(\omega)
  \vert}\mu^{\prime\prime}(\omega) \right ]
\vert {\bf  E}_0 \vert^2 e^{-2 {\bf k}^{\prime\prime} \cdot {\bf r} }
\ . 
\end{equation}

\noindent
Already at this point we can notice that the coefficient in the
parentheses in the above formula appears to be somewhat strange.
Indeed, if $\mu$ is purely imaginary (e.g., near a resonance), this
coefficient becomes $\vert \epsilon \vert + \epsilon^{\prime\prime}$.
If, in addition, $\vert \epsilon^{\prime} \vert \gg
\epsilon^{\prime\prime}$, the heating rate becomes proportional to
$\vert \epsilon^{\prime} \vert$.

We note that a propagating wave in an absorbing infinite medium grows
exponentially in the direction $-{\bf k}^{\prime\prime}$. To avoid the
unbounded growth, one has to consider a half space $z>0$ into which an
incident wave enters, e.g., from vacuum, and apply the condition
$\hat{\bf z} \cdot {\bf k}^{\prime\prime} >0$. However, a wave which
is refracted from vacuum into an absorbing medium is necessarily
evanescent. This follows immediately from the fact that the projection
of the wave vector on the plane $z=0$ must be continuous at the
interface and, therefore, purely real (since it is real in vacuum). In
the case of evanescent waves, the equalities ${\bf k} \cdot {\bf k}^*
= \vert \mu\epsilon \vert (\omega/c)^2$ and ${\bf k} \cdot {\bf E}_0^*
= 0$ do not hold and the expression for $q^{\rm (conv)}$ becomes more
complicated. This effect is not important for weakly absorbing media
and it will not be discussed here. We only note that the expression
(\ref{Q2}) which will be obtained below from the definition $q =
\langle {\bf J} \cdot {\bf E} \rangle$ applies to both running and
evanescent waves and, in any case, differs from
(\ref{Q_Landau_monochrom_planewave}) or (\ref{Q1}).

\subsection{Second derivation of $q$}
\label{subsec:second}

We now compute the same quantity as in the previous subsection but
using a different, presumably equivalent, definition. Namely, we write

\begin{equation}
\label{Q_work}
q = \langle {\bf J} \cdot {\bf E} \rangle \ ,
\end{equation}

\noindent
where ${\bf J}$ is the total current in the medium induced by the
propagating electromagnetic field. We again emphasize that this
current is formed by the charged particles (bound and conduction
electrons, ions, etc.)  which make up the medium. Equation
(\ref{Q_work}) is simply the mathematical formulation of the statement
that, in a stationary state, the heating rate is equal to the
(time-averaged) work that the electric field exerts on the medium per
unit time per unit volume.

We again consider a plane monochromatic wave with the electric field
given by the first equation in (\ref{EH_real_complex}). The current
also has the form of a plane wave:

\begin{equation}
\label{J_real_complex}
{\bf J} = {\rm Re}[{\bf J}_0 e^{i({\bf k} \cdot {\bf r} - \omega t)}]
\ .
\end{equation}

\noindent
Note that the above formula is valid only inside the medium volume. At
the boundary, there is an additional surface current related to
magnetization. This current and the corresponding contribution to the
heating rate will be considered separately in
Section~\ref{sec:surface}. We now focus on the volume contribution to
the heating rate and denote the corresponding quantity by $q^{(V)}$.
Time-averaging results in

\begin{equation}
\label{Q_JE_av}
q^{(V)} = \frac{1}{2} {\rm Re} ({\bf J}_0 \cdot {\bf E}_0^*) e^{-2 {\bf
    k}^{\prime\prime} \cdot {\bf r}} \ . 
\end{equation}

\noindent
To find ${\bf J}_0$, we write the two curl Maxwell equations as

\begin{equation}
\label{Maxwell}
c \nabla \times {\bf E} = - \partial {\bf B}/\partial t \ , \ \ c\nabla
\times {\bf B} = \partial {\bf E} / \partial t + 4\pi {\bf J}  \ ,
\end{equation}

\noindent
where ${\bf J}$ is given by (\ref{J_def}). Note that the above
equations are equivalent to the usual macroscopic Maxwell equations if
we define the auxiliary fields ${\bf D} = {\bf E} + 4\pi {\bf P}$ and
${\bf H} = {\bf B} - 4\pi {\bf M}$ and use (\ref{J_def}). By taking
the time derivative of the second equation in (\ref{Maxwell}) and
substituting $\partial {\bf B} / \partial t$ from the first equation,
we find that

\begin{equation}
\label{J_E}
- 4\pi \partial {\bf J}/ \partial t = c^2 \nabla \times \nabla \times
{\bf E} + \partial^2 {\bf E} / \partial t^2 \ .
\end{equation}

\noindent
At the next step, we substitute (\ref{J_real_complex}) and the first
equation in (\ref{EH_real_complex}) into (\ref{J_E}) to obtain

\begin{equation}
\label{J_0_def}
{\bf J}_0 = - (c^2 /4\pi i \omega) [{\bf k}
\times {\bf k} \times {\bf E}_0 + (\omega/c)^2 {\bf E}_0] \ .
\end{equation}

\noindent
We then use ${\bf k} \times {\bf k} \times {\bf E}_0 = - ({\bf k}
\cdot {\bf k}) {\bf E}_0$ and ${\bf k} \cdot {\bf k} = \mu\epsilon
(\omega/c)^2$ (this holds for both propagating and evanescent waves)
to further simplify the above expression for ${\bf J}_0$, which
becomes

\begin{equation}
\label{J_0_em}
{\bf J}_0 = \frac{\omega}{4\pi i} [\mu(\omega) \epsilon(\omega) - 1]
{\bf E}_0 \ .
\end{equation}

\noindent
Upon substitution of the above expression into (\ref{Q_JE_av}), we
arrive at

\begin{equation}
\label{Q2}
q^{(V)} = \frac{\omega \vert {\bf E}_0 \vert^2 }{8\pi} {\rm Im}\left[ \mu(\omega)
  \epsilon(\omega) \right] e^{-2 {\bf k}^{\prime\prime} \cdot {\bf r}} \ .
\end{equation}

\noindent
I shall generalize this result to the case of monochromatic field
${\bf E} = {\rm Re} [{\bf E}_{\omega}({\bf r})\exp(-i\omega t)]$ (not
necessarily a plane wave) in Eq.~(\ref{Q_monochrom}) below.

The expression (\ref{Q2}) must be compared to (\ref{Q1}). The
respective formulae obviously differ. The reason for this discrepancy
and the correct choice of the expression for $q$ are discussed in
Section~\ref{sec:Poynting}.

\subsection{The two expressions for the heating rate and the constraints on
  $\epsilon$ and $\mu$ that follow from them}

If we accept the conventional result for the heating rate as correct,
the second law of thermodynamics requires that, in a passive medium,
$q^{\rm (conv)} > 0$, where $q^{\rm (conv)}$ is given for plane waves
by (\ref{Q_Landau_monochrom_planewave}) or (\ref{Q1}). If, however, we
assume that the alternative expression (\ref{Q2}) is correct, then the
second law requires that the volume contribution to the heating rate
$q^{(V)}$ be positive (proof is given in Section~\ref{sec:thermo}).
From this, a different constraint on the possible values of $\epsilon$
and $\mu$ is obtained.

We note right away that for the conventional expression
(\ref{Q_Landau_monochrom_planewave}) to be positive, it is not {\em
  necessary} that

\begin{equation}
\label{eps_mu_im}
\epsilon^{\prime\prime} > 0 \ , \ \ \ \mu^{\prime\prime} > 0 \ ,
\end{equation}

\noindent
although the above inequality is a {\em sufficient} condition. The
{\em sufficient and necessary condition} is

\begin{equation}
\label{eps_mu_im_1}
\vert \mu \vert \epsilon^{\prime\prime} + \vert \epsilon \vert
\mu^{\prime\prime} > 0 \ . 
\end{equation}

\noindent
Neither (\ref{eps_mu_im}) nor (\ref{eps_mu_im_1}) prohibit negative
refraction. We note that it can be argued (assuming
(\ref{Q_Landau_monochrom_planewave}) is correct) that {\em both}
inequalities $\epsilon^{\prime\prime} > 0$ and $\mu^{\prime\prime} >
0$ must hold simultaneously and independently to guarantee positivity
of the heating rate~\cite{depine_04_1,efros_04_1}. Indeed, the
condition (\ref{eps_mu_im_1}) was derived for a plane wave and is,
therefore, not the most general.

The alternative expression (\ref{Q2}) imposes a different constraint
on $\epsilon$ and $\mu$. For $q^{(V)}$ given by the expression
(\ref{Q2}) to be positive, the {\em sufficient and necessary}
condition is

\begin{equation}
\label{Q2_positivity}
\epsilon^{\prime} \mu^{\prime\prime} + \mu^{\prime}
\epsilon^{\prime\prime} > 0 \ .
\end{equation}

\noindent
Thus, $\epsilon^{\prime}$ and $\mu^{\prime}$ can not be simultaneously
negative while $\epsilon^{\prime\prime}$ and $\mu^{\prime\prime}$ are
positive. In particular, (\ref{Q2_positivity}) prohibits negative
refraction in the sense that if the wave number $k$ satisfies $k^2 =
\epsilon\mu (\omega/c)^2$, its real and imaginary parts have the same
sign, independently of the choice of the square root branch.

Finally, note the following interesting fact. If we put $\mu=1$,
formulae (\ref{Q1}) and (\ref{Q2}) become identical. But if we put
$\epsilon=1$ (e.g., in a purely magnetic material), the two
expressions still differ. This is suggestive of the fact that magnetic
losses are not properly accounted for in one of these formulae.

\section{Correct Expressions for the Poynting Vector and the Heating Rate}
\label{sec:Poynting}

Both formulae (\ref{Q1}) and (\ref{Q2}) do not obviously contradict
any of the basic physical principles, such as the conservation laws.
Therefore, we must choose the correct expression for $q$ on less
fundamental grounds. To this end, we examine the origin of the two
definitions (\ref{Q_Landau}) and (\ref{Q_work}).

The expression (\ref{Q_Landau}) is obtained from $q =- \langle \nabla
\cdot {\bf S} \rangle$ where the Poynting vector ${\bf S}$ is given by
Eq.~(\ref{S_H}). Differentiation leads to

\begin{equation}
q = -\frac{c}{4\pi} \langle {\bf H} \cdot (\nabla \times {\bf E}) -
{\bf E} \cdot (\nabla \times {\bf H}) \rangle \ . 
\end{equation}

\noindent
One then uses the macroscopic Maxwell equations to express $\nabla
\times {\bf E}$ and $\nabla \times {\bf H}$ in terms of the
corresponding time derivatives to arrive at (\ref{Q_Landau}). In the
stationary case, when there is no accumulation of electromagnetic
energy anywhere inside the medium, the formula $q =- \langle \nabla
\cdot {\bf S} \rangle$ is undoubtedly correct. It is a mathematical
expression of the statement that the total electromagnetic energy
which enters into a small volume $\delta V$ through its surface is
entirely consumed to compensate for the irreversible (absorptive)
losses in that volume.

The alternative formula, (\ref{Q_work}), is also a first-principles
definition of the absorbed power per unit volume and appears to be
unassailable. I emphasize again that the quantity ${\bf J}$ in
(\ref{Q_work}) is the {\em total internal current} produced by all
charged particles that compose the material. This includes bound
electrons, conductivity electrons (if such are present), ions in the
case of plasmas, etc.

So far, it appears that in either of the two approaches, the only
formula that can be doubted is the definition of the Poynting vector
(\ref{S_H}) which was used to derive (\ref{Q_Landau}). It should be
noted that in standard textbook expositions, the form the Poynting
vector is postulated rather than derived. Thus, for example, Schwinger
{\em et al.} (in {\em Classical
  Electrodynamics}~\cite[\$7.1]{schwinger_book_98}) consider the identity

\begin{equation}
\label{identity}
\frac{c}{4\pi} \nabla \cdot ({\bf E} \times {\bf H}) + \frac{1}{4\pi}
\left( {\bf E} \cdot \frac{\partial {\bf D}}{\partial t} + {\bf H} \cdot
  \frac{\partial {\bf B}}{ \partial t} \right ) = 0
\end{equation}

\noindent
which is trivially obtainable from the macroscopic Maxwell equations
in the absence of external currents. Then Schwinger {\em et al.}
write: ``Our aim is to write this result as a local energy
conservation law.  We immediately identify, from the divergence term,
the energy flux or Poynting vector ${\bf S}$ to be ${\bf S}= (c/4\pi)
{\bf E} \times {\bf H}$.'' The argument is, however, mathematically
flawed. Indeed, one can take any scalar function $f({\bf r})\neq 0$
whose integral over the body volume is zero and write it as a
divergence of a vector field, $f({\bf r}) = \nabla \cdot {\bf F}({\bf
  r})$, where ${\bf F}({\bf r})$ vanishes outside of the body.  One
then can add $(c/4\pi)\nabla \cdot {\bf F}$ to the first term in the
left-hand side of (\ref{identity}) and subtract it from the second
term, and the identity will still hold.  According to the logic of
Refs.~\onlinecite{schwinger_book_98}, one then has to define the
Poynting vector as ${\bf S} = (c/4\pi) [{\bf E} \times {\bf H} + {\bf
  F}]$. Note that the field ${\bf F}$ does not need to be solenoidal,
so that not only the definition of ${\bf S}$ is changed, but also of
its divergence. This ambiguity in the conventional definition of ${\bf
  S}$ leads to a substantial strain. To quote Schwinger again, ``...
More intractable is the identification of the last term in
(\ref{identity}).''  I argue that such identification is, indeed,
intractable because the term in question has no physical meaning.

A somewhat different approach to deriving the conventional expression
for ${\bf S}$ is adopted by Landau and Lifshitz in {\em
  Electrodynamics of Continuous Medium}~\cite[\$80]{landau_ess_84}.
First, it is shown that Eq.~(\ref{S_H}) is valid in non-magnetic media
where ${\bf H} = {\bf B}$. Then Landau and Lifshitz argue that the
normal component of ${\bf S}$ should be continuous when a wave crosses
an interface between two media. Since the tangential components of
both ${\bf E}$ and ${\bf H}$ are continuous, the normal component of
${\bf S}$ defined by (\ref{S_H}) is continuous as well. Therefore,
(\ref{S_H}) should be valid in any media, including those with
dispersion and a magnetic response.

I do not dispute here that the tangential components of ${\bf E}$ and
${\bf H}$ are continuous as long as there is no surface current at the
interface which is formed by charges which are {\em external to the
  medium}. Note that sometimes such currents are referred to as ``free
currents'' (even though they do not include the current of free
electrons in the case of conductors). However, I claim that continuity
of the normal component of ${\bf S}$ is an incorrect boundary
condition for interfaces that separate two media at least one of which
is magnetic. Indeed, it is known~\cite[\$29]{landau_ess_84} that
nonzero magnetization creates surface currents which are restricted to
a very thin layer near the medium boundary. These currents are formed
by the charges of the medium and, therefore, do not cause the
tangential component of ${\bf H}$ to become discontinuous as follows
immediately from the equation $c\nabla \times {\bf H} = \partial {\bf
  D} /\partial t$. When a wave crosses an interface in which such
surface current is flowing, a finite fraction of its energy is lost to
the (positive or negative) work exerted by the electric field on the
surface current. In this case, the normal component of ${\bf S}$
experiences a discontinuity. The role of the surface currents and
their input to the heating rate is discussed in
Section~\ref{sec:surface} below.

To obtain the correct expression for the Poynting vector, we start
with the microscopic electric and magnetic fields, ${\bf e}$ and ${\bf
  h}$. The spatial averages of these fields are~\cite[\$1 and
\$29]{landau_ess_84} $\overline{\bf e} = {\bf E}$ and $\overline{\bf
  h} = {\bf B}$. Here the bar denotes spatial averaging over
physically small volumes.  Further, we write ${\bf e} = {\bf E} +
\delta {\bf e}$ and ${\bf h} = {\bf B} + \delta{\bf h}$, where
$\delta{\bf e}$ and $\delta{\bf h}$ are the fluctuating parts of the
fields. The microscopic expression for the Poynting vector is

\begin{equation}
\label{S_micro}
{\bf s} = \frac{c}{4\pi} {\bf e} \times {\bf h} \ .
\end{equation}

\noindent
We now average the above expression as follows:

\begin{equation}
{\bf S} \equiv \overline{\bf s} = \frac{c}{4\pi}\left[ \ {\bf E} \times
  {\bf B} + \overline{\delta{\bf e}}\times {\bf B} + {\bf E} \times
  \overline{\delta{\bf h}} + \overline{\delta{\bf e} \times \delta{\bf
      h}} \ \right] \ .
\end{equation}

\noindent
By definition, $\overline{\delta{\bf e}} = \overline{\delta{\bf h}} =
0$. The term $\overline{\delta{\bf e} \times \delta{\bf h}}$ is
quadratic in field fluctuations and can be omitted as small. It should
be also noted that in materials which are random but isotropic on
average, this term is identically zero by symmetry.  We thus arrive at
the expression (\ref{S_B}) for the Poynting vector.  We note that in
vacuum, the Poynting vector is expressed in terms of the electric and
magnetic fields. There is no conceivable physical reason why this
should change if the field propagates through a material medium. But
the average value of the magnetic field in the medium is ${\bf B}$,
not ${\bf H}$. Despite the fact that ${\bf H}$ is commonly called the
"magnetic field", it is actually an auxiliary quantity.

Let us adopt the definition (\ref{S_B}) for the Poynting vector and
compute $q^{(V)}$ from $q^{(V)} =- \langle \nabla \cdot {\bf S}
\rangle$ for a general monochromatic fields of the form

\begin{eqnarray}
\label{E_D_mono}
{\bf E} = {\rm Re}\left[ {\bf E}_\omega({\bf r}) e^{-i\omega t}
\right] \ ,\
  {\bf D} = {\rm Re}\left[ {\bf D}_\omega({\bf r}) e^{-i\omega t}
\right] \ , \ \ \\ 
\label{H_B_mono}
{\bf H} = {\rm Re}\left[ {\bf H}_\omega({\bf r}) e^{-i\omega t} 
\right] \ , \ 
   {\bf B} = {\rm Re}\left[ {\bf B}_\omega({\bf r}) e^{-i\omega t}
\right] \ , \ \
\end{eqnarray}

\noindent
where ${\bf D}_\omega = \epsilon(\omega) {\bf E}_\omega$, ${\bf
  B}_\omega = \mu(\omega) {\bf H}_\omega$, $c\nabla \times {\bf
  E}_\omega = i\omega {\bf B}_\omega$ and $c\nabla \times {\bf
  H}_\omega = -i\omega {\bf D}_\omega$. At the moment, we do not
consider the heating rate at the surface where ${\bf S}$ has a
discontinuity. We then have:

\begin{equation}
\langle {\bf S} \rangle = \frac{c}{8\pi} {\bf E}_\omega^* \times {\bf
  B}_\omega 
\end{equation}

\noindent
and

\begin{eqnarray}
q^{(V)} &&= -\nabla \cdot \langle {\bf S} \rangle \nonumber \\
  &&= \frac{c}{8\pi} {\rm Re}\left[-\nabla \cdot ({\bf E}_\omega^*
    \times {\bf B}_\omega) \right] \nonumber \\
  &&= \frac{c}{8\pi} {\rm Re}\left[ {\bf E}_\omega^* \cdot (\nabla \times
    {\bf B}_\omega) - {\bf B}_\omega \cdot (\nabla
    \times {\bf E}_\omega^*) \right] \nonumber \\
  &&=\frac{c}{8\pi} {\rm Re}\left[{\bf E}^*_\omega \cdot
    (\nabla\times \mu(\omega) {\bf H}_\omega) + \frac{i\omega}{c}{\bf
      B}_\omega \cdot {\bf B}_\omega^*\right] \nonumber \\ 
  &&=\frac{c}{8\pi} {\rm Re}\left[ \frac{-i\omega}{c} \mu(\omega) {\bf
      E}_\omega^* \cdot {\bf D}_\omega \right] \nonumber \\
  &&=\frac{\omega \vert {\bf E}_\omega \vert^2}{8\pi}{\rm
    Im}[\mu(\omega)\epsilon(\omega)] \ . 
\label{Q_divS}
\end{eqnarray}

\noindent
We thus have derived the following formula for $q^{(V)}$:

\begin{equation}
\label{Q_monochrom}
q^{(V)} = \frac{\omega \vert {\bf E}_\omega \vert^2}{8\pi}{\rm
    Im}[\mu(\omega)\epsilon(\omega)] \ . 
\end{equation}

\noindent
If we set ${\bf E}_\omega = {\bf E}_0 \exp(i {\bf k} \cdot {\bf r})$,
the above expression coincides with formula (\ref{Q2}) which was
derived previously from the definition $q^{(V)} = \langle {\bf J}
\cdot {\bf E} \rangle$ for the case of a plane wave of the form
(\ref{EH_real_complex}).

Next, consider the definition $q^{(V)} = \langle {\bf J} \cdot {\bf E}
\rangle$. We have already used this definition to compute $q^{(V)}$
for a plane wave with the result given by Eq.~(\ref{Q2}). Now we
repeat the calculation for more general monochromatic fields
(\ref{E_D_mono}),(\ref{H_B_mono}). The current (except at the medium
surface) can also be written in a similar form, namely,

\begin{equation}
\label{J_mono}
{\bf J} = {\rm Re}\left[ {\bf J}_\omega({\bf r}) e^{-i\omega t}
\right] \ , 
\end{equation}

\noindent
where 

\begin{equation}
\label{J_mono_def}
{\bf J}_\omega = \frac{1}{4\pi} \left( c\nabla\times {\bf B}_\omega + i\omega
{\bf E}_\omega \right) \ .
\end{equation}

\noindent
We then write

\begin{eqnarray}
q^{(V)} &&= \langle {\bf J} \cdot {\bf E} \rangle = \frac{1}{2} {\rm
  Re}({\bf J}_\omega \cdot {\bf E}_\omega^*) \nonumber \\
  &&=\frac{1}{8\pi} {\rm Re}\left[i\omega {\bf E}_\omega \cdot {\bf
      E}_\omega^* + c (\nabla \times {\bf B}_\omega) \cdot {\bf
      E}_\omega^* \right] \nonumber \\
  &&=\frac{c}{8\pi} {\rm Re}\left[ \mu(\omega) (\nabla \times {\bf
      H}_\omega) \cdot {\bf E}_\omega^* \right] \nonumber \\
  &&=\frac{c}{8\pi} {\rm Re}\left[
    \frac{-i\omega}{c}\mu(\omega)\epsilon(\omega) {\bf E}_\omega \cdot
  {\bf E}_\omega^* \right] \nonumber \\
  &&=\frac{\omega \vert {\bf E}_\omega \vert^2}{8\pi}{\rm
    Im}[\mu(\omega)\epsilon(\omega)] \ . 
\label{Q_JE}
\end{eqnarray}

\noindent
The result of this calculation coincides with (\ref{Q_monochrom}).

Thus, we can conclude that if ${\bf S}$ is defined by (\ref{S_B}), the
two definition of the heating rate, $q^{(V)} =-\langle \nabla \cdot {\bf S}
\rangle$ and $q^{(V)} = \langle {\bf J} \cdot {\bf E} \rangle$ are
equivalent and we have the statement of local energy conservation
which, in the stationary case, reads

\begin{equation}
\label{energy_conservation}
\langle {\bf J}\cdot {\bf E} \rangle + \langle \nabla \cdot {\bf S}
\rangle = 0 \ .
\end{equation}

\section{Heating Rate at the Surface and the Total Absorbed Heat}
\label{sec:surface}

The derivation of the heating rate form the formula (\ref{Q_work}) was
so far restricted to points inside the medium. In this section, the
additional surface term $q^{(S)}$ is derived.

First, consider the definition (\ref{Q_work}). In the monochromatic
case, the current in this formula is given by (\ref{J_mono_def}) where
${\bf B}_\omega = \mu(\omega) {\bf H}_\omega$. In the sequence of
equalities (\ref{Q_JE}), I have, at one point, replaced the term
$\nabla \times {\bf B}_\omega$ by $\mu(\omega) \nabla \times {\bf
  H}_\omega$.  This operation is only valid inside the medium volume.
Close to the surface, we must write

\begin{equation}
\label{msc_1}
\nabla \times {\bf B}_\omega = \nabla \times \mu({\bf r}) {\bf
  H}_\omega = \mu({\bf r})
\nabla \times {\bf H}_\omega + [\nabla \mu({\bf r})] \times {\bf H}_\omega \ ,
\end{equation}

\noindent
where the dependence of $\mu$ on position has been indicated
explicitly. In the case of macroscopically homogeneous media, $\nabla
\mu({\bf r}) = 0$ everywhere except at the surface, where $\mu({\bf
  r})$ experiences a discontinuity. If we restrict attention to points
${\bf r}$ which are on the surface, evaluation of $\nabla \mu({\bf
  r})$ results in the additional surface
current~\cite[\$29]{landau_ess_84}

\begin{equation}
\label{J_s_differ}
{\bf J}_\omega^{(S)} = -c \hat{\bf n} \times {\bf M}_\omega \ ,
\end{equation}

\noindent
where $\hat{\bf n}$ is the outward unit normal to the boundary at the
point ${\bf r}$. Note that the definition of the surface current
(\ref{J_s_differ}) does not contain a spatial delta-function and that
${\bf J}_\omega^{(S)}$ has different physical units then the volume
current ${\bf J}$.

We now find the surface contribution to the heating rate as

\begin{equation}
\label{q_s_JE_1}
q^{(S)} = \frac{1}{2}{\rm Re} \left( {\bf J}_\omega^{(S)} \cdot {\bf
  E}_\omega^* \right) \ .
\end{equation}

\noindent
A straightforward derivation results in 

\begin{equation}
\label{q_s_JE_2}
q^{(S)} = \frac{c}{8\pi}{\rm Re} \left[(1 - \mu) (\hat{\bf n} \times
  {\bf H}_\omega) \cdot {\bf E}_\omega^* \right] \ .
\end{equation}

\noindent
It is also possible to start from the definition $q = - \nabla \cdot
{\bf S}$, take into account the fact that the normal component of
${\bf S}$ defined by (\ref{S_B}) experiences a discontinuity at the
medium boundary, and arrive, in a straightforward manner at 

\begin{equation}
\label{q_s_JE_3}
q^{(S)} = \frac{c}{8\pi}{\rm Re} \left[(1 - \mu) ({\bf
    H}_\omega \times {\bf E}_\omega^*) \cdot \hat{\bf n} \right] \ .
\end{equation}

\noindent
Since ${\bf a} \cdot ({\bf b} \times {\bf c}) = {\bf b} \cdot ({\bf c}
\times {\bf a})$, the two expressions (\ref{q_s_JE_2}) and
(\ref{q_s_JE_3}) are identical.

The total heat absorbed by the body, $Q$, is given by
Eq.~(\ref{W_def}). It is easy to see that this quantity is the same as
in the conventional theory. Indeed, $Q$ can be computed by integrating
the energy flux through any surface enclosing the body. Such surface
can be drawn in free space where the conventional expression for the
Poynting vector (\ref{S_H}) and the expression derived in this paper
(\ref{S_B}) coincide. Therefore, the proposed change in the form of
the Poynting vector and of the heating rate does not affect any of the
previously established results for differential or integral cross
sections, such as the Mie formulae for extinction, absorption and
scattering cross sections of spheres.

The surface contribution to the heating rate derived in this section
requires several additional comments. The obvious distinction between
the surface term $q^{(S)}$ (\ref{q_s_JE_3}) and the volume term
$q^{(V)}$ (\ref{Q_monochrom}) is that the volume term is proportional
to the frequency $\omega$ while the surface term is not. Of course, it
is incorrect to say that $q^{(V)}$ always vanishes in the
zero-frequency limit because $\lim_{\omega \rightarrow 0} [\omega
\epsilon(\omega) \mu(\omega)] = 4\pi i\sigma \mu(\omega=0)$, where
$\sigma$ is the static conductivity of the material~\cite{fn1}.
However, it appears that the surface term does not vanish in the
zero-frequency limit even if we formally set $\sigma=0$.  This
possibility is worrisome and is discussed below.

First, this paper is primarily concerned with the high-frequency
superficial magnetism which originates due to the loop-like
conductivity currents flowing in elementary cells of composite
materials. The magnetic susceptibility of such composites identically
vanishes in the zero-frequency
limit~\cite{pendry_99_1,podolskiy_03_1}. Moreover, the formula
(\ref{Q_work}) in which the current is given by (\ref{J_def}) assumes
that all currents obey the classical laws of motion. This may not be
the case when magnetization is caused by spin aligning, as in the
cases of ferro- and para-magnetism. However, even in the case of
ferromagnetism, the current $c\nabla \times {\bf M}$ is a macroscopic
quantity. According to the Ehrenfest theorem, all macroscopic
quantities obey the classical laws of motion. Nevertheless, it should
be acknowledged that some of the phenomena associated with
ferromagnetism, such as the hysteresis, are clearly outside of the
theoretical frame of the classical electrodynamics of continuous
media. Such effects can be accounted for phenomenologically but not in
a fully self-consistent way.

Thus, the zero-frequency limit might not be the proper test for the
theory developed in this paper. I will, however, argue that it is
possible to apply this theory to the zero-frequency limit without
obtaining unphysical effects or contradictions. To this end, I
consider below two simple examples.

The first example is a straight ferromagnetic or paramagnetic
cylindrical wire of radius $a$, conductivity $\sigma$ and permeability
$\mu$ carrying a current of uniform density $J$ directed along the
axis of the wire. I disregard here the Hall effect that results in a
non-uniform current distribution over the wire cross
section~\cite{assis_99_1}. The quantities $\mu$ and $\sigma$ are
purely real at zero frequency. Then my theory predicts that the volume
will be heated at the rate (per length $L$ of the wire)

\begin{equation}
\label{Q_wire_V}
Q^{(V)}/L = \pi a^2 \sigma \mu E^2
\end{equation}

\noindent
and the surface will be heated or cooled at the rate

\begin{equation}
\label{Q_wire_S}
Q^{(S)}/L = \pi a^2 \sigma (1-\mu) E^2 \ .
\end{equation}

\noindent
In a ferro- and paramagnetic materials, $\mu>1$ (for the case of
ferromagnetic, the nonlinearity of the magnetization curve and the
magnetic memory of the material must be taken into account, which is
not a trivial task), so that the surface term is negative.  But the
total heat produced in the system is the sum of both contributions,
namely,

\begin{equation}
\label{Q_wire_Tot}
Q/L = Q^{(V)}/L + Q^{(S)}/L = \pi a^2 \sigma E^2 \ ,
\end{equation}

\noindent
which is the Joule's law. In a steady state, the overall flux of
thermal energy through the wire surface, which is the experimentally
measurable quantity (e.g., in a calorimeter) is given by $Q/L$, in
agreement with the Joule's law.

The second example is a magnetic object placed in crossed external
electric and magnetic fields. For simplicity, consider a long cylinder
uniformly magnetized along its axis. The magnetization will create
loop-like surface currents that flow around the cylinder axis. We now
place the cylinder in an external electric field which is
perpendicular to the cylinder axis (this example has been previously
considered by Pershan~\cite{pershan_63_1}). If the cylinder is
conducting (as are most ferromagnets), the tangential component of the
electric field at the cylinder surface, as well as the electric field
inside the cylinder, vanish and we obtain $q^{(S)} = q^{(V)} = 0$, as
expected. A somewhat more complicated situation arises if we formally
set $\sigma=0$ and $\mu-1 \neq 0$. The volume term $q^{(V)}$ is still
zero in this case, but the surface term $q^{(S)}(\varphi)$ may become
locally nonzero (here $\rho,z,\varphi$ are the cylindrical
coordinates). Even though it can be easily seen that

\begin{equation}
\label{Q_cylinder_int}
\int_0^{2\pi} q^{(S)}(\varphi)d\varphi = 0 \ ,
\end{equation}

\noindent
we still expect no local heating or cooling of the surface in the
static equilibrium. The contradiction is resolved by noting that the
state of the cylinder described above can not be its true state of
equilibrium and that the initial assumption $\sigma=0$ and $\mu-1\neq
0$ was unphysical. This assumption contradicts the mechanical
equilibrium of charges that make up the circular magnetization
currents. Classically, these charges rotate with constant angular
velocity around the cylinder axis due to a phenomenological radial
force. It is, however, not possible to introduce a phenomenological
restoring {\em tangential} force such as the harmonic restoring force
in the Lorentz model of dielectrics.  Indeed, such restoring
tangential force will preclude the magnetization current from flowing
in the first place.  Consequently, imposition of an external
tangential force (due to the external electric field) in the absence
of a tangential restoring force is bound to break the equilibrium of
the system.  Specifically, the external electric field will cause
electric charge to accumulate on the cylinder surface until the
tangential component of the electric field is completely nullified.
The resultant state will be the true static equilibrium of the system.
The conclusion is that magnetized objects can not have identically
zero conductivity. Of course, the value of $\sigma$ can be small, but
so is usually the value of $\mu-1$. Another important consideration is
that, in addition to conductivity, magnetics also have some dielectric
response whose effect is to diminish the tangential electric field at
the body surface.

The example considered above suggests that the conventional definition
of the Poynting vector (\ref{S_H}) is erroneous because it predicts
existence of an equilibrium state which contradicts mechanical
stability of the system.

\section{Thermodynamic Considerations and Impossibility of Negative
  Refraction}
\label{sec:thermo}

Many authors believe that the unique properties of the negative
refraction materials originate from the fact that the phase velocity
and the Poynting vector in such media are oppositely directed. This
property is sometimes referred to as ``backward propagation''. For
example, to quote Marques {\em at al.}\cite[\$1.2]{marques_book_08},
``\ldots most of the surprising unique electromagnetic properties of
these media arise from this backward propagation property.'' If the
expression (\ref{S_B}) for the Poynting vector is correct, as I argue
in this paper, then the phase velocity and the Poynting vector always
point in the same direction and ``backward propagation'' is
impossible, at least in electromagnetically homogeneous media. I will,
however, apply more fundamental thermodynamic considerations to show
that the inequality (\ref{NR_cond}) is physically prohibited,
regardless of whether it results in those ``surprising unique
effects'' or not.

To this end, it is instructive to introduce the "accessible heat".
This is the heat (either positive or negative) which can be
transferred from the body to a heat reservoir on a time scale which is
short compared to the time scales associated with heat diffusion in
the body. Obviously, this is the heat generated at the surface. Let

\begin{equation}
\label{W_S_def}
Q^{(S)} \equiv \oint_S q^{(S)}({\bf r})d^2 r = Q_+^{(S)} - Q_-^{(S)} \
.
\end{equation}

\noindent
Here $Q_+^{(S)}$ is obtained by integration over the surface areas
where $q^{(S)}({\bf r})$ is positive and $Q_-^{(S)}$ is obtained by
integration over the surface areas where $q^{(S)}({\bf r})$ is
negative. Let us further assume that the material exhibits negative
refraction and $q^{(V)}({\bf r})$ is negative, so the heat generated
in the volume,

\begin{equation}
\label{W_V_def}
Q^{(V)} \equiv \int_V q^{(V)}({\bf r})d^3 r
\end{equation}

\noindent
is also negative. Then we have 

\begin{equation}
\label{Q_QS}
Q = -\vert Q^{(V)}\vert + Q_+^{(S)} - Q_-^{(S)} 
\end{equation}

\noindent
or, equivalently, 

\begin{equation}
\label{QS_Q}
Q_+^{(S)} = Q + \vert Q^{(V)}\vert + Q_-^{(S)} > Q \ .
\end{equation}

\begin{figure}[h]
\centerline{\psfig{file=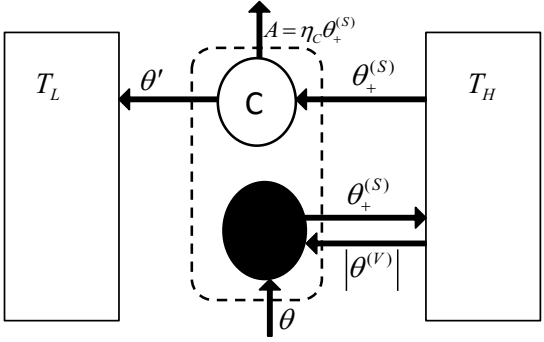,width=8.3cm,bbllx=500bp,bblly=300bp,bburx=720bp,bbury=500bp,clip=t}}
\caption{\label{fig:1} A cyclic process involving a
  negative refractive index material that violates the Carnot theorem.
  The black oval represents a negative refraction medium and the white
  oval is an ideal Carnot engine.}
\end{figure}

\noindent
Thus, the positive accessible heat is greater than the total heat
absorbed in the body. I will now demonstrate that this contradicts the
Carnot theorem and, moreover, can be used to create a {\em perpetuum
  mobile} of the second kind. To see that this is, indeed the case,
consider the cyclic process shown in the Fig.~\ref{fig:1}.

In this cycle, the following events happen: (1) A negative-refraction
sample represented by the black oval (referred to as the ``body''
below) at the initial temperature $T_H$ is irradiated for a period of
time $\Delta t$ which is short compared to the time scales associated
with heat diffusion in the body, yet long compared to the
electromagnetic oscillations period, so that radiation is almost
monochromatic. The body absorbs the energy $\theta = Q \Delta t$ from
the radiation field. (2) The body is brought in contact with a heat
reservoir at the temperature $T_H$ which has very high heat
conductivity; the amount of thermal energy $\theta_+^{(S)} =
Q_+^{(S)}\Delta t$ generated at the body's surface is transferred
adiabatically to this reservoir. (3) The body is disconnected from the
reservoir and heat diffusion takes place in the body until the
equilibrium temperature $T^\prime < T_H$ is reached. (4) An ideal
Carnot engine is operated for one cycle between the hot reservoir and
a colder reservoir whose temperature is $T_L<T_H$. The Carnot engine
absorbs the heat $\theta_+^{(S)}$ from the hot reservoir, makes useful
work $A = \eta_C \theta_+^{(S)}$ and rejects some amount of heat
$\theta^\prime = \theta_+^{(S)} - A$ to the cold reservoir.  Here

\begin{equation}
\label{eta_C}
\eta_C = 1 - T_L/T_H 
\end{equation}

\noindent
is the efficiency of an ideal Carnot engine.  (5) The body is again
brought in contact with the hot reservoir; now the heat $\vert
\theta^{(V)} \vert$ flows back from the hot reservoir to the body. In
the end of this process, the body has the temperature $T_H$.  (6) We
disconnect the body from the hot reservoir. Now the cycle is complete
and the system has returned to its original state.  Note that steps
(1) and (2), as well as (3) and (4) can be combined, i.e., occur
simultaneously.

The net effect of the above thermodynamic transformation is the
following: The electromagnetic field has done the work $\theta$ on the
body which was immediately dissipated into heat $\theta$; we then
converted this heat into the useful work $A$. The overall efficiency
of this process is

\begin{equation}
\label{eta_A}
\eta = \frac{A}{\theta} = \eta_C \frac{\theta_+^{(S)}}{\theta_+^{(S)} -
  \vert \theta^{(V)} \vert} = \eta_C \frac{1}{1 -
  \vert \theta^{(V)} \vert / \theta_+^{(S)}} > \eta_C \ 
\end{equation}

\noindent
in violation of the Carnot theorem. Moreover, we can operate a {\em
  perpetuum mobile} of the second kind if $A > \theta$ or,
equivalently, $\eta > 1$. This is achieved if $T_L / T_H < \vert
\theta^{(V)} \vert / \theta_+^{(S)}$.  There is no physical reason why
this condition can not be met. In particular, it can be met quite
easily in the case of low-loss negative refraction materials such that
$\vert \theta^{(V)} \vert / \theta_+^{(S)} = 1 - \delta$ where $\delta
\ll 1$. Then even relatively small temperature difference $T_H - T_L$
would be sufficient to extract more energy from the heat reservoir
than was absorbed from the electromagnetic field. Note that in order
to obtain the contradiction, it is essential that the ``accessible''
heat $\theta_+^{(S)}$ be larger than the total absorbed heat $\theta$.
This is always the case for negative refraction materials, as was
shown above. Therefore, I conclude that negative refraction is
impossible.

It is also possible to use negative refraction to construct a
refrigeration cycle in violation of the Carnot theorem.

\section{Heating Rate and Causality}
\label{sec:causality}

As is well known, the causality principle requires that
$\epsilon(\omega)$ and $\mu(\omega)$ have no singularities in the
upper half-plane when viewed as functions of the complex variable
$\omega$. In the lower complex half-plane, these functions may have
singularities. We will assume now that all such singularities are
simple poles. If, in addition, we account for the symmetry property
$\epsilon(-\omega) = \epsilon^*(\omega)$ and $\mu(-\omega) =
\mu^*(\omega)$, we can write $\epsilon$ and $\mu$ in the most general
form as

\begin{equation}
\label{eps_mu_chi}
\epsilon(\omega) = 1 + 4\pi \chi_e(\omega) \ , \ \ \mu(\omega) = 1 +
4\pi \chi_m(\omega) \ , 
\end{equation}

\noindent
where

\begin{eqnarray}
\label{chi_e}
\chi_e(\omega) = \sum_k \frac{f_k(\omega)}{a_k^2 - \omega^2 - i\alpha_k \omega} \
, \\
\label{chi_m}
\chi_m(\omega) = \sum_k \frac{g_k(\omega)}{b_k^2 - \omega^2 - i\beta_k \omega} \
.
\end{eqnarray}

\noindent
Here $\chi_e$ and $\chi_m$ are the electric and magnetic
susceptibilities, respectively, $a_k, b_k, \alpha_k$ and $\beta_k$
are coefficients and $f_k(\omega)$ and $g_k(\omega)$ are analytical
functions of the frequency which have no singularities and satisfy
$f_k(-\omega) = f_k^*(\omega)$ and analogously for $g_k$. The
representation (\ref{eps_mu_chi})-(\ref{chi_m}) is customary in the
theory of dispersion. In the electric case, the functions $f_k$ are
usually positive constants interpreted as oscillator strengths. In the
magnetic case, the typical form of $g_k(\omega)$ is $g_k(\omega) =
c_k\omega^2$ where the coefficients $c_k$ can be negative, i.e., in
the case of diamagnetics. Terms proportional to $\omega^{2n}$ with
$n>1$ in the Taylor expansion of $f_k(\omega)$ and $g_k(\omega)$ are
not usually considered because of the physical requirement that both
$\epsilon$ and $\mu$ are bounded when $\omega\rightarrow \infty$.
The physical interpretation of the remaining constants appearing in
(\ref{chi_e}),(\ref{chi_m}) is as follows: $a_k$ and $b_k$ are the
resonance frequencies and $\alpha_k$ and $\beta_k$ are the respective
relaxation constants.

We wish to examine whether the representation
(\ref{eps_mu_chi})-(\ref{chi_m}) is compatible with the inequality
$q(\omega) \propto {\rm Im}[\epsilon(\omega) \mu(\omega)] > 0$ which
must hold for all positive frequencies. Obviously, the latter imposes
some constraints on the coefficients appearing in formulae
(\ref{chi_e})-(\ref{chi_m}). However, it is easy to show that
coefficients that satisfy $q(\omega)>0$ for all $\omega >0$ do exist.
For instance, a sufficient condition for $q>0$ is obtained when one of
the susceptibilities is significantly smaller than the other.  Let us
write

\begin{equation}
{\rm Im}(\epsilon\mu) = 4\pi {\rm
  Im}(\chi_e + \chi_m) + (4\pi)^2 {\rm Im}(\chi_e \chi_m) \ .
\end{equation}

\noindent
Since $(4\pi)^2 {\rm Im} (\chi_e \chi_m) > - (4\pi)^2 \vert \chi_ e
\chi_m \vert$ and ${\rm Im}(\chi_e + \chi_m) > {\rm Im}(\chi_e)$, a
sufficient condition for ${\rm Im}(\mu\epsilon) > 0$ is $\vert \chi_m
\vert < \chi_e^{\prime\prime} /4\pi \vert \chi_e \vert$. This
inequality can always be satisfied in the whole frequency range for
sufficiently small coefficient $c_k$ (assuming
$g_k(\omega)=c_k\omega^2$). Analogously, $q>0$ if $\vert \chi_e \vert
< \chi_m^{\prime\prime} /4\pi \vert \chi_m \vert$.

Thus, we have obtained two sufficient conditions for $q>0$. However,
these conditions are not necessary.  To derive a condition which is
both sufficient and necessary, consider the special case of a single
electric and single magnetic resonance with positive and
frequency-independent oscillator strengths, $f_e^2$ and $g_e^2$. That
is, assume that $\epsilon$ and $\mu$ are given by

\begin{eqnarray}
\label{eps_1r}
\epsilon(\omega) = 1 + 4\pi \frac{f_e^2}{\omega_e^2 - \omega^2 -
  i\gamma_e \omega} \ , \\
\label{mu_1r}
\mu(\omega) = 1 + 4\pi \frac{f_m^2}{\omega_m^2 - \omega^2 - i\gamma_m
  \omega} \ ,
\end{eqnarray}

\noindent
where all coefficients are positive. Then a straightforward
calculation shows that

\begin{equation}
\label{Im_z}
\frac{A_e(\omega) A_m(\omega)}{4\pi\omega} {\rm Im}(\mu\epsilon) = a
\omega^4 + b\omega^2 + c \ ,
\end{equation}

\noindent
where

\begin{eqnarray}
\label{a_def}
a = && f_e^2 \gamma_e + f_m^2 \gamma_m \ , \\
b = && f_m^2\gamma_m(\gamma_e^2 - 2\omega_e^2) +
       f_e^2\gamma_e(\gamma_m^2 - 2\omega_m^2) \nonumber \\
    && \hspace{2.5cm} - 4\pi f_e^2 f_m^2 (\gamma_e + \gamma_m) , \\ 
\label{b_def}
c = && 4\pi f_e^2 f_m^2 (\gamma_m \omega_e^2 + \gamma_e\omega_m^2) 
\nonumber \\
    && \hspace{2.5cm} + f_m^2\gamma_m\omega_e^4 + f_e^2\gamma_e\omega_m^4
\label{c_def}
\end{eqnarray}

\noindent
and $A_e(\omega), A_m(\omega)$ are positive factors defined by

\begin{eqnarray}
&& A_e(\omega) = (\omega_e^2 - \omega^2)^2 + (\gamma_e \omega)^2 \ , \\
&& A_m(\omega) = (\omega_m^2 - \omega^2)^2 + (\gamma_m \omega)^2 \ .
\end{eqnarray}

\noindent
The right-hand side of (\ref{Im_z}) is a quadratic polynomial in
$\omega^2$ with a positive free term $c$. If we assume that relaxation
constants are small so that $\gamma_e^2 < 2\omega_e^2$ and $\gamma_m^2
< 2\omega_m^2$, the coefficient $b$ is negative. In this case, the
necessary and sufficient condition that the polynomial is positive is
${\mathcal D} = b^2 - 4ac <0$. A tedious but straightforward
calculation yields the following expression for the discriminant
${\mathcal D}$:

\begin{eqnarray}
\frac{\mathcal D}{\gamma_e \gamma_m} = && \gamma_e
\gamma_m\left[f_m^4(\gamma_e^2 - 4\omega_e^2) + f_e^4(\gamma_m^2 -
  4\omega_m^2)\right] \nonumber \\
&& + 2f_e^2 f_m^2\left[\gamma_e^2 \gamma_m^2 - 2(\gamma_m^2 \omega_e^2 +
  \gamma_e^2 \omega_m^2)  \right. \nonumber \\
&& \hspace{1.5cm} - \left. 2(\omega_e^4 + \omega_m^4) +
  4\omega_e^2\omega_m^2 \right] \nonumber \\ 
&& -8\pi f_e^2 f_m^2 \left\{ f_m^2\left[ \gamma_e^2 + \gamma_e\gamma_m
    + 2(\omega_m^2 - \omega_e^2)\right] \right. \nonumber \\
&& \hspace{1.5cm} + \left. f_e^2\left[ \gamma_m^2 + \gamma_m\gamma_e
    + 2(\omega_e^2 - \omega_m^2)\right] \right\} \nonumber \\
&& + (4\pi)^2 f_e^4 f_m^4 (\gamma_e + \gamma_m)^2 \ .
\label{D_def}
\end{eqnarray}

\noindent
Assuming that $b<0$, the sufficient and necessary condition for ${\rm
  Im}(\mu\epsilon) > 0$ is that the above expression is negative.
This, of course, leads to a very complicated inequality. However, if
the relaxation constants are small compared to all other physical
scales of the problem, this inequality can be simplified and reads

\begin{eqnarray}
&& (\omega_e^2 - \omega_m^2)^2 + 4\pi(\omega_m^2 - \omega_e^2)(f_m^2 -
f_e^2) \nonumber \\
&& \hspace{2cm} > 4\pi^2 f_e^2f_m^2 \frac{(\gamma_e + \gamma_m)^2}{\gamma_e\gamma_m} \ .
\label{ineq}
\end{eqnarray}

\begin{figure}[h]
\centerline{\psfig{file=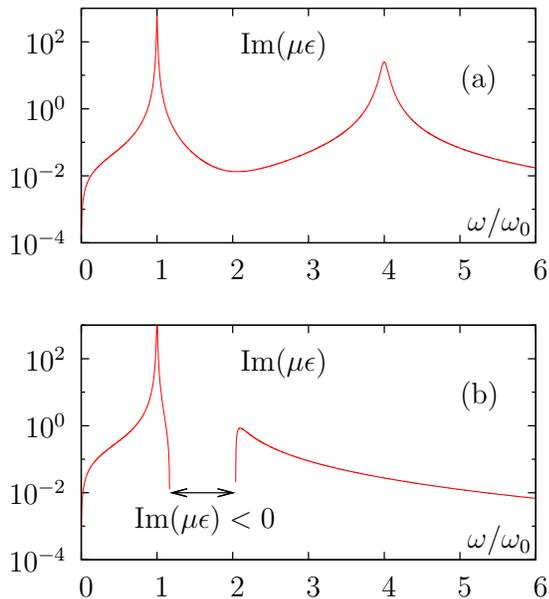,width=8.3cm,bbllx=200bp,bblly=475bp,bburx=450bp,bbury=750bp,clip=}}
\caption{\label{fig:2} Plots of ${\rm Im}[\mu(\omega)\epsilon(\omega)]$ where
  $\epsilon$ and $\mu$ are given by (\ref{eps_1r}),(\ref{mu_1r}) for
  different sets of parameters.  Plot (a): $\omega_m=\omega_0$,
  $\omega_e=4\omega_0$, $f_m=0.5\omega_0$, $f_e=\omega_0$, $\gamma_m =
  0.01\omega_0$ and $\gamma_e=0.1\omega_0$; $\omega_0$ is an arbitrary
  frequency scale.  The parameters satisfy inequality (\ref{ineq}).
  Plot (b): same parameters as in plot (a) but $\omega_e=2\omega_0$.
  With this change, (\ref{ineq}) is no longer satisfied in the
  frequency range denoted by the horizontal arrow. Negative values of
  ${\rm Im}[\mu(\omega)\epsilon(\omega)]$ are not shown in the plots
  due to the use of logarithmic scale.}
\end{figure}

To illustrate how inequality (\ref{ineq}) works, I have plotted ${\rm
  Im}(\mu\epsilon)$ as a function of $\omega$ for two different sets
of parameters (see Fig.~\ref{fig:2} caption for details). As can be
seen, if the parameters satisfy (\ref{ineq}), ${\rm Im}(\mu\epsilon) >
0$ in the whole frequency range (Fig.~\ref{fig:2}a). If, however, we
use a set of parameters that does not satisfy (\ref{ineq}), there
appears a frequency range in which ${\rm Im}(\mu\epsilon) < 0$
(Fig.~\ref{fig:2}b). As I have argued above, $q \propto {\rm
  Im}(\mu\epsilon)$.  Therefore, the set of parameters used to compute
the curve shown in Fig.~\ref{fig:2}b leads to negative heating rate in
the frequency range indicated by the horizontal arrow. These values of
parameters violate the second law of thermodynamics and, therefore,
can not be realized in any material, either natural or artificial.

Thus, we have seen that the expression for the heating rate derived in
this paper does not contradict causality. However, it imposes
constraints on the possible values of constants in the dispersion
formulae (\ref{chi_e}),(\ref{chi_m}). Generally, these constraints are
very complicated mathematically. In the simplest case of one magnetic
and one electric resonance ($\epsilon$ and $\mu$ given by formulae
(\ref{eps_1r}) and (\ref{mu_1r})), the condition is that the
discriminant (\ref{D_def}) is negative. In the limit of small
relaxation constants $\gamma_e$ and $\gamma_m$, this condition can be
approximated by the much more simple inequality (\ref{ineq}).

We finally note that the condition (\ref{ineq}) is only sufficient but
not necessary if the relaxation is so strong that the coefficient $b$
given by formula (\ref{b_def}) is positive. Also, in the case when one
of the resonance frequencies is zero, as is the case for electric 
permittivity of conductors, inequality (\ref{ineq}) may become an
inaccurate approximation of the more general inequality ${\mathcal D}
<0$ where ${\mathcal D}$ is given by (\ref{D_def}).

\section{Heating Rate in Anisotropic and Nonlocal Media}
\label{sec:anisotropy}

The general case of a medium with magnetic and electric anisotropy and
nonlocality is quite complicated. The wave vector ${\bf k}$ of a plane
wave propagating in such a medium can be found from the following
condition:

\begin{equation}
\label{k_det}
{\rm det}\left\vert \hat{\epsilon}^{-1} {\bf k} \times \hat{\mu}^{-1}
  {\bf k} \times \  + \left(\frac{\omega}{c}\right)^2 \right\vert = 0
\end{equation}

\noindent
where $\hat{\epsilon} = \hat{\epsilon}(\omega,{\bf k})$ and $\hat{\mu}
= \hat{\mu}(\omega,{\bf k})$ are ${\bf k}$-dependent tensors. 

The dispersion relation (\ref{k_det}) is simplified for the case of
propagating waves. A {\em propagating} (as opposed to an {\em
  evanescent}) wave is characterized by a wave vector ${\bf k} = k
\hat{\bf u}$ where $k$ is a complex scalar and $\hat{\bf u}$ is a
purely real unit vector such that $\hat{\bf u} \cdot \hat{\bf u} = 1$.
Thus, a propagating wave can, in principle, experience spatial decay.
The important point is that, in the propagating case, the wave vector
is completely characterized by a direction in space (the unit vector
$\hat{\bf u}$) and by a single scalar $k$. We can utilize this
property to rewrite (\ref{k_det}) as

\begin{equation}
\label{k_det_prop}
{\rm det} \left\vert - k^2 \hat{T} + \left(\frac{\omega}{c}\right)^2
\right\vert = 0 \ ,
\end{equation}

\noindent
where $\hat{T} = - \hat{\epsilon}^{-1} \hat{\bf u} \times
\hat{\mu}^{-1} \hat{\bf u} \times$. In general, the $3\times 3$ tensor
$\hat{\bf T}$ is symmetric but not Hermitian. Therefore, its
eigenvectors and eigenvalues, denoted here by ${\bf v}_j$ and
$1/\tau_j$, are complex. For each direction $\hat{\bf u}$, the wave
number of a propagating wave is determined from one of the equations
$k^2 = \tau_j (\omega/c)^2$ while the polarization of the $j$-th mode
is given by ${\bf E}_0 = a{\bf v}_j$, $a$ being an arbitrary complex
constant.

Interestingly, it is possible to make a statement about the
restrictions that are imposed by the condition $q^{(V)}>0$ on the wave
number $k$ without explicitly solving the dispersion equations.  We
note that the formula $q = \langle {\bf J} \cdot {\bf E} \rangle$ is
valid in any electromagnetically homogeneous media. We then consider
monochromatic, propagating plane wave with the wave vector ${\bf k}$,
so that the fields are of the form

\begin{eqnarray}
&& {\bf E} = {\rm Re}\left[ {\bf E}_0 e^{i({\bf k} \cdot {\bf r} -
    \omega t)} \right] \ , \\
&& {\bf B} = {\rm Re}\left[ {\bf B}_0 e^{i({\bf k} \cdot {\bf r} -
    \omega t)}
\right] \ , \\
&& {\bf J} = {\rm Re}\left[ {\bf J}_0 e^{i({\bf k} \cdot {\bf r} -
    \omega t)}
\right] \ ,
\end{eqnarray}

\noindent
and

\begin{equation}
4\pi {\bf J}_0 = i\omega {\bf E}_0 + i c {\bf k} \times {\bf
  B}_0 \ . 
\end{equation}

\noindent
We then obtain

\begin{equation}
\label{Q_a}
q^{(V)} = \frac{c e^{-2 {\bf k}^{\prime\prime} \cdot {\bf r}}}{8\pi} {\rm Im}
  \left[ ({\bf k} \times {\bf B}_0) \cdot {\bf E}_0^* \right ] \ .
\end{equation}

\noindent
We now use ${\bf k} \times {\bf B}_0 = (c/\omega) {\bf k} \times {\bf
  k} \times {\bf E}_0$ and the identity ${\bf a} \times {\bf b} \times
{\bf c} = {\bf b} ({\bf a} \cdot {\bf c}) - {\bf c} ({\bf a} \cdot
{\bf b})$ to arrive at the following result:

\begin{equation}
\label{Q_b}
q^{(V)} = \frac{\omega e^{-2 {\bf k}^{\prime\prime} \cdot {\bf r}}}{8\pi(\omega/c)^2}
{\rm Im} \left[ \vert {\bf E}_0 \vert^2 ({\bf
    k} \cdot {\bf k}) - ({\bf k} \cdot {\bf E}_0) ({\bf k} \cdot {\bf
    E}_0^*) \right]  \ .
\end{equation}

\noindent
Note that we have not used any constitutive relations in the
derivation of (\ref{Q_b}). Also note that the wave vector ${\bf k}$
must satisfy the dispersion relation (\ref{k_det}).  If ${\bf k} =
k\hat{\bf u}$, it is always possible to write $({\bf k} \cdot {\bf
  E}_0) ({\bf k} \cdot {\bf E}_0^*) = \cos^2\theta \vert {\bf E}_0
\vert^2$, where $\theta$ is a purely real angle. From this, we obtain
the final expression for $q$:

\begin{equation}
\label{Q_c}
q^{(V)} = \frac{\omega \vert {\bf E}_0 \vert^2 e^{-2 {\bf k}^{\prime\prime}
    \cdot {\bf r}}}{8\pi} \frac{\sin^2 \theta \ {\rm Im} (k^2)}{(\omega/c)^2} \ .
\end{equation}

The phenomenon of negative refraction requires that the direction in
which a wave exponentially decays due to absorption in the medium is
opposite to its phase velocity. Mathematically, this means that the
real and imaginary parts of the complex wave number must have opposite
signs. But for this to be true, it is required that ${\rm Im} (k^2) <
0$. However, Eq.~(\ref{Q_c}) implies that ${\rm Im} (k^2) > 0$. Thus,
negative refraction is not physically attainable even in anisotropic
and nonlocal media.

Finally, we note that one can formally choose the polarization and the
wave number in such a way that $\sin^2\theta = 0$ so that the medium
does not absorb radiation. However, it is easy to see that waves with
$\sin^2\theta = 0$ do not satisfy the dispersion relation
(\ref{k_det}).

\section{Discussion}
\label{sec:discussion}

In this section, I address certain anticipated objections to the
theory developed in this paper, as well as discuss some of its
limitations.

The first and the most obvious objection is that there have been a
number of works which claim experimental demonstration of negative
refraction in electromagnetically homogeneous materials. Such
experiments can be classified into two kinds. The direct-kind
experiments measure the deflection of a beam passing through an
experimental sample made of a subwavelength-structured
``metamaterial''. However, in most experiments of this kind, the
linear size of the smallest metamaterial element, $\ell$ is not much
smaller than even the vacuum wavelength $\lambda$. Strictly speaking,
$\ell$ should be compared to the wavelength {\em inside} the material.
Further, the more physically relevant parameter is $k\ell = 2\pi
\ell/\lambda$. In typical experimental demonstrations of negative
refraction, this parameter is of the order of unity. Under these
circumstances, interpretation of the experimental results in terms of
the bulk constants $\epsilon$ and $\mu$ is problematic. There are also
experiments in which the negative refraction is measured indirectly by
means of measuring the transmission and reflection coefficients $t$
and $r$ of a subwavelength-structured thin film. Here, as in the case
of direct-kind experiments, it is very difficult to achieve $k\ell \ll
1$.  Additionally, the indirect-kind experiments rely on phase
measurements of high-frequency electromagnetic fields and on a
analytical procedure of extracting $\epsilon$ and $\mu$ from the
measurements of $t$ and $r$. Both of these tasks are notoriously
difficult and have recently been subject to some
controversy~\cite{kildishev_07_1,grigorenko_07_1,simovski_07_1}.

The second objection is based on the factually incorrect, yet
widespread belief that the magnetic field can, under certain
circumstances, do work, i.e., on magnetic moments. Theory developed in
this paper is based on the premise that only electric field can do
work.  The question of whether the magnetic force can do work is
simultaneously simple and complicated. Of course, it immediately
follows from the expression for the Lorentz force that the magnetic
force does no work on a moving charged particle. Yet, apart from this
simple observation which can be found in most textbooks on classical
electrodynamics, there has been almost no serious discussion of this
question in scientific literature. At the same time, situations in
which the magnetic force is {\em apparently} doing work are quite
abundant. Recently, the question was addressed in a mathematically
rigorous way by Deissler~\cite{deissler_08_1}. In this reference, it
is shown that the magnetic force does no work on a classical magnetic
moment under any circumstances. It is further shown that the magnetic
force does no work on an atom where magnetization is due to orbital
angular momentum.  Finally, Deissler shows that there is a fully
self-consistent description of the quantum spin in which the magnetic
field does no work either.

Third objection is based on the belief that composite media can be
assigned {\em effective} medium parameters which describe
(approximately) some phenomena associated with wave propagation
through such media but not the others. I believe that such contention
was expressed, for example, by Simovski~\cite{simovski_07_1}, although
implicitly. My reply to this is that most experimentally measurable
quantities, such as the intensity, are bilinear in the electric and
magnetic fields. Therefore, any useful homogenization model must
correctly predict such quadratic combinations, including the Poynting
vector and the heating rate.

Fourth objection is that the ``magnetic'' current $c\nabla \times {\bf
  M}$ is somehow different in its physical properties from the
``electric'' current $\partial {\bf P} / \partial t$ and, therefore,
obeys different laws of motion. Of course, in the case of
metamaterials, both currents have exactly the same physical origin.
But even in the most general case, both currents are macroscopic and
there is no valid physical basis to apply different laws of motion to
them. It is also not possible to do so mathematically.  Assume that we
know a vector field ${\bf J}({\bf r})$.  Assume also that we know that
${\bf J}({\bf r}) = {\bf J}_e({\bf r}) + {\bf J}_m({\bf r})$ where
${\bf J}_e({\bf r}) = \partial {\bf P}({\bf r}) / \partial t$ and
${\bf J}_m({\bf r})= c \nabla \times {\bf M}({\bf r})$. Is it possible
to find uniquely ${\bf J}_e$ and ${\bf J}_m$ if we know ${\bf J}$ (but
not ${\bf P}$ or ${\bf M}$)? It is known that ${\bf J}_m$ is
solenoidal.  If it were also known that ${\bf J}_e$ is irrotational,
we would be able to use the Helmholtz theorem to uniquely decompose
${\bf J}$ into the irrotational and the solenoidal parts corresponding
to ${\bf J}_e$ and ${\bf J}_m$. But the only instance when ${\bf J}_e$
is irrotational is the static case.  Therefore, beyond strict statics,
there is no unique way to disentangle the term $c \nabla \times {\bf
  M}$ from the total current ${\bf J}$.

The fifth set of objections is related to the zero-frequency limit.
This limit is discussed in detail in Section~\ref{sec:surface}. Here I
would like to reiterate the following. The theory developed in this
paper is based on the fundamental assumption that all currents obey
the same classical laws of motion. Although the author sees no
physical reason for this assumption to be untrue even in the zero
frequency limit, reasonable caution must be exercised when applying
the results to magnetization caused by quantum spins. It is
theoretically possible that the magnetic susceptibility which is due
to spin alignment has different physical and mathematical properties
when compared to the magnetic susceptibility which is due to classical
eddy currents. In this case, the total permeability must be written as
$\mu -1  = (\mu_{\rm classical}-1) + (\mu_{\rm quantum} - 1)$, where $\mu_{\rm
  quantum}$ is the contribution to the total permeability due to
quantum effects which are manifest only at low frequencies, and $\mu$
should be replaced by $\mu_{\rm classical}$ in the expressions for the
heating rate derived in this paper.

Finally, the essential requirement for applicability of the results
derived in this paper is that the medium is electromagnetically
homogeneous or can be effectively homogenized. Mathematically, this
means that the medium must support running plane waves as its
electromagnetic modes. This condition is not satisfied in photonic
crystals and similar structures which support propagating modes in the
form of {\em Bloch waves}.

\section{Conclusions}
\label{sec:conclusions}

The article has the following conclusions: (i) The correct definition
of the Poynting vector in magnetic media is (\ref{S_B}). If this
definition is used, then the local energy conservation law in the form
(\ref{energy_conservation}) holds, where ${\bf J}$ is given by
(\ref{J_def}). (ii) The heating rate $q^{(V)}$, defined as the amount
of energy converted to heat per unit time per unit volume, is
proportional to the factor ${\rm Im}(\mu\epsilon)$ {\em inside the
  volume occupied by the material}; heating rate at the surface is
given by Eqs.~(\ref{q_s_JE_2}) or (\ref{q_s_JE_3}) in
Section~\ref{sec:surface}. (iii) It follows from the detailed
thermodynamic considerations of Section~\ref{sec:thermo} that negative
refraction contradicts the second law of thermodynamics. This
statement holds for active or passive media and in the presence of
anisotropy and spatial dispersion, as is shown in
Section~\ref{sec:anisotropy}.

\bibliographystyle{unsrt} 
\bibliography{abbrev,master,book,local}

\begin{thebibliography}{10}
\bibitem{fn1}
In the zero-frequency limit, we must replace ${\bf E}_\omega$ by $2{\bf E}$,
  where ${\bf E}$ is the real-valued electric field, and similarly for the
  magnetic field.
\end{thebibliography}

\end{document}